\documentclass[%
 preprint,
 amsmath,amssymb,
 aps,floatfix,
]{revtex4-2}

\usepackage{amsmath}
\usepackage{amssymb}

\usepackage{subfig}
\usepackage{graphicx}%
\usepackage{dcolumn}%
\usepackage{bm}%
\usepackage{booktabs}
\usepackage{multirow}

\newcommand{\eat}[1]{}

\begin{document}

\preprint{APS/123-QED}

\title{
Game-Theoretic Analysis of Adversarial Decision Making \\
in a Complex Sociophysical System
}

\author{Andrew C. Cullen}
 \email{andrew.cullen@unimelb.edu.au}
\affiliation{%
 School of Computing and Information Systems, University of Melbourne, Parkville, Australia
}%

\author{Tansu Alpcan}
\affiliation{
Department of Electronic and Electrical Engineering, University of Melbourne, Parkville, Australia
}%
\author{Alexander C. Kalloniatis}
\affiliation{%
DST Group, Canberra
}%

\date{\today}%

\begin{abstract}

We apply Game Theory to a mathematical representation of two competing teams of agents connected within a complex network, where the ability of each side to manoeuvre their resource and degrade that of the other depends on their ability to internally synchronise decision-making while out-pacing the other. Such a representation of an adversarial socio-physical system has application in a range of business, sporting, and military contexts. Specifically, we unite here two physics-based models, that of Kuramoto to represent decision-making cycles, and an adaptation of a multi-species Lotka-Volterra system for the resource competition. For complex networks we employ variations of the Barab\'asi-Alberts scale-free graph, varying how resources are initially distributed between graph hub and periphery. We adapt as equilibrium solution Nash Dominant Game Pruning as a means of efficiently exploring the dynamical decision tree. Across various scenarios we find Nash solutions where the side initially concentrating resources in the periphery can sustain competition to achieve victory except when asymmetries exist between the two. When structural advantage is limited we find that agility in how the victor stays ahead of decision-state of the other becomes critical.
\end{abstract}

\maketitle

\section{Introduction}

Physics-based modelling of competitive
social-systems has been a consistent
thread in complexity research.
\cite{schelling1958strategy, raquel2007application, berner2019dota, vinyals2019grandmaster, halevy2008team, karabiyik2020understanding, golestani2021game}. 
Such systems involve
not only a competition of resources
between two teams,
with one seeking to have more or degrade
those of the other, but also decision-making, where strategies are
selected to achieve best position
in the resource fight. Game Theory, where both sides
select strategies within concepts
such as the Nash equilibrium,
is a natural extension of such models.
In this paper we develop this 
where the topology of network structures becomes a factor in one 
social-system, organisation or team
achieving advantage over the other.

The model we employ has been
developed over
a number of years. Firstly,
a cognitive process is
represented through 
coupled Kuramoto oscillators \cite{Kuramoto1984}, which can usefully represent the Boyd Observe-Orient-Decide-Act (OODA) model of decision making~\cite{BoydOODA, OSINGA2013603} between competing individuals or organisations~\cite{demazy2018game, zuparic2021adversarial}. 
Next, the resource competition
between organisations is represented
by a variation of the Lotka-Volterra
multi-species predator-prey model,
for example the Lanchester
model ~\cite{Lanchester1916, mackay2006lanchester, jin2011modeling}
generalised to a network of resources
of two adversarial teams, `Blue' and `Red'
\cite{Kalloniatis2020NetLanch}.
These models may be united
into a single representation
of two organisations, where each seeking to degrade
the resource of the other, enabled by optimised inta-organisational collective decision-making to gain advantage
\cite{Ahern2020,kondakov2022dynamics}.
Our unique contribution
is a time-dependent game theoretic analysis of such a model under a Stackleberg Security-Strategy framework~\cite{yin2010stackelberg}. 
This intersection of sociophysical models and game theory may
test concepts for 'victory' used in various contexts, for example the value of holding resources in reserve during armed conflict ('Economy of Force')~\cite{skinner1988}; or whether competing business firms should establish a niche or seek market dominance~\cite{Moorthy2014}.
We examine how
initial resource apportionment depends on network topology,
albeit in a single stylised use-case, 
in terms of solutions representing
Nash strategies; deviation from these yields no gain for winner or loser.

In this paper Sect.~\ref{sec:model} summarises networked, adversarial `Boyd--Kuramoto' model, with the game theoretic tools used to construct solutions described in Sect.~\ref{sec:numerics}. Our main results are then presented in Sect.~\ref{sec:results}, which provide `heat-maps' describing the transitions in competitive outcomes for different structural resource
distribution. We conclude with a brief summary and discussion in Sect.~\ref{sec:conclusion}

\section{Model}\label{sec:model}

The specific variant of the decision-making dynamics that we employ is the extended `Networked Boyd--Kuramoto--Lanchester' (NBKL) model~\cite{demazy2018game, Kalloniatis2020NetLanch, Ahern2020}, which represents a Blue team (whose members are of the set $\mathcal{B}$) and Red team (of $\mathcal{R}$) adversarially by way of 
\begin{equation}\label{e:nbkl_positioning}
\frac{d \theta_i}{d t} = \mathcal{H}(p_{i}) \left( \omega_i - \sum_{j \in \mathcal{R} \cup \mathcal{B}} \mathcal{H}(p_{i}) \mathcal{K}_{ij} \sigma_{ij} \sin \left(\theta_i - \theta_j - \Phi_{ij}\right) \right)\enspace.  \\
\end{equation}
For each agent $i \in \mathcal{R} \cup \mathcal{B}$ the phase $\theta_i$ describes their state within a decision-making cycle (eg Observe-Orient-Decide-Act), while $\mathcal{H}(p_i)$ is a Heaviside function that
disables an agent when its resource
degrades to zero: it ceases both to engage in the decision-making process and to apply degradation on an adversary. An agent left to itself advances decisions with frequency $\omega_i$, however this varies
with the number of links with other agents
in the network through the adjacency matrix
\begin{equation}
    \mathcal{K} = \epsilon + \mathcal{M}\enspace.
\end{equation}
The matrix elements $\epsilon_{ij}$, $\mathcal{M}_{ij}$ are zero, except where nodes $i$ and $j$ are linked, either between or within their teams, respectively for $\epsilon$ and $\mathcal{M}$. The matrix $\sigma = \zeta \epsilon$ then describes the strength of coupling between any two agents. Finally, $\Phi_{ij} \in [0, \pi]$ represents an offset in the decision-state between two nodes; 
this uses the formalism of `frustration' in magnetic systems but here
represents an intended advantage one agent seeks in decision-making relative to another. This systems nonlinear dynamics ensure that neither team is guaranteed to achieve their selected state as a stable solution; chaotic dynamics may deny one side a constant advantage ahead of the other.

The evolution of the resource dynamics is described by
\begin{align}\label{eqn:nbkl_populations}
&\frac{d p_i}{d t} =  \mathcal{H}(p_{i}) \left(  \sum_{h \in \mathcal{R} \cup \mathcal{B}} \mathcal{H}(p_h) \mathcal{M}_{ih}\frac{\Gamma_{i,h} + \Gamma_{h,i}}{2} \left(\delta_h p_h - \delta_i p_i\right) \cdot
\right. \nonumber\\
&\quad \frac{\cos\left(\theta_h  - \theta_i\right) + 1}{2} - \sum_{k \in \mathcal{R} \cup \mathcal{B}}\mathcal{H}(p_k) \epsilon_{ik}\kappa_{ik}p_{k}d_{k} \cdot\\
&\qquad \left. \frac{\sin(\theta_k - \theta_i ) + 1}{2}O_k \right) \nonumber \enspace.
\end{align}
This captures each team's internal ability to redistribute resources 
through synchronised decision-making
in the first sum; and the 
ability to degrade the other team's resources through decision advantage in the second sum. Within this, the parameters $\Gamma = \gamma \mathcal{M}$ and $\kappa = \kappa^{RB} \epsilon$ respectively represent the rates of resupply and degradation that can be achieved through network links. 
For simplicity, the same networks
represent communication paths for
decision-synchronisation and resource manoeuvre-paths for each team. 
These flows are moderated by the local synchronisation `order parameter'
\cite{Kuramoto1984}, $\mathcal{O}_k$ and the parameters $\delta_k$ and $d_k$, which respectively take the form
\begin{align}
    O_k &= \frac{\left| \sum_{m \in \mathcal{M}} \mathcal{M}_{km} \mathcal{H}(p_m) \exp^{\sqrt{-1} \theta_m} + \epsilon_2 \right|}{\sum_{m \in \mathcal{M}} \mathcal{M}_{km} \mathcal{H}(p_m) + \epsilon_2} \nonumber \\
    \delta_k &= \frac{1}{\sum_{m \in \varepsilon} \mathcal{\varepsilon}_{km} p_m + 1} \nonumber \\
    d_k &= \frac{1}{\sum_{m \in \varepsilon} \mathcal{\varepsilon}_{km} \mathcal{H}(p_m) + \epsilon_2}\enspace. \nonumber
\end{align}
Thus, local coherence of decision-making
enhances such manoeuvre. 
The nodes at which engagement between the two teams occur (where resource degradation is applied) is called the
adversarial surface between the two communications and resource flow networks of each team.
A summary of the parameter space of these quantities can be seen within Table~\ref{tab:parameter_summary}. The specific choice of parameters has been motivated by previous works \cite{Ahern2020, zuparic2023friend}, with specific emphasis on constructing equally capable teams for matched topologies, and to ensure that the speed of decision-making cycles and coupling is an order of magnitude faster than the speed of
degradation of resources. 

\begin{table*}
\caption{Summary of Parameters, where $B$ and $R$ represent the sizes of the $\mathcal{B}$ and $\mathcal{R}$ teams.%
}
\label{tab:parameter_summary}
  \centering
\scalebox{1.00}{  
  \begin{tabular}{lll}
    \toprule
Parameter & Description & Type \\
\cmidrule(r){1-1} \cmidrule(r){2-2} \cmidrule(r){3-3}
$\theta_i$ & Phase (Agent Decision State) & $\mathbb{R}^{B + R}$ \\
$p_i$ & Agent Strength & $\mathbb{R}^{B + R}$ \\
$\mathcal{M}$ & Intra-Team Adjacency Matrices & $\mathbb{R}^{(R+B)^2}$ \\
$\gamma$ & Intra-Team Network Coupling & $\mathbb{R}$ \\
$\epsilon$ & Inter-Team Adjacency Matrices & $\mathbb{R}^{(R+B)^2}$ \\
$\zeta$ & Inter-Team Network Coupling & $\mathbb{R}$ \\
$\mathcal{K}$ & Phase coupling matrix  & $\mathbb{R}^{(R+B)^2}$ \\
$\Phi_{ij}$ & Agent Positioning & $[0, \pi]^{B + R}$ \\
$\delta$ & Inhibitor of internal force transfer & $\mathbb{R}^{B + R}$ \\
$d$ & Inhibitor of adversarial outcomes & $\mathbb{R}^{B + R}$ \\
$\omega_i$ & Decision Speed & $\mathbb{R}^{B + R} = 0.1$ \\
$O_i$ & local order parameter & $\mathbb{R}^{B + R}$ \\
$\mathcal{H}(\cdot)$ & Heaviside function & $\{0, 1\}$\\
$\epsilon_1, \epsilon_2$ & Scaling parameters & $10^{-20}$, $10^{-20}$ \\
$\zeta$ & Coupling constant & $0.5$\\
$\kappa^{RB}$ & Attritional coupling constant & $0.05$ \\
    \bottomrule
  \end{tabular}
  }
\end{table*}

\subsection{Network Topology}
\begin{figure*}[htbp!]
   \subfloat[$k = 1$]{%
      \includegraphics[width=0.3\textwidth]{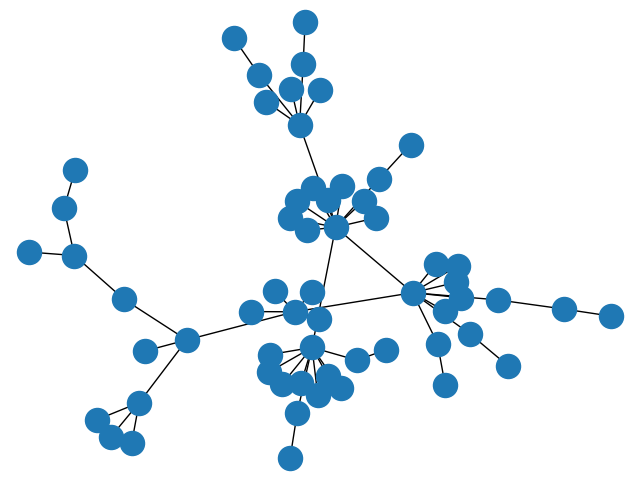}}
\quad \quad
   \subfloat[$k = 2$]{%
      \includegraphics[width=0.3\textwidth]{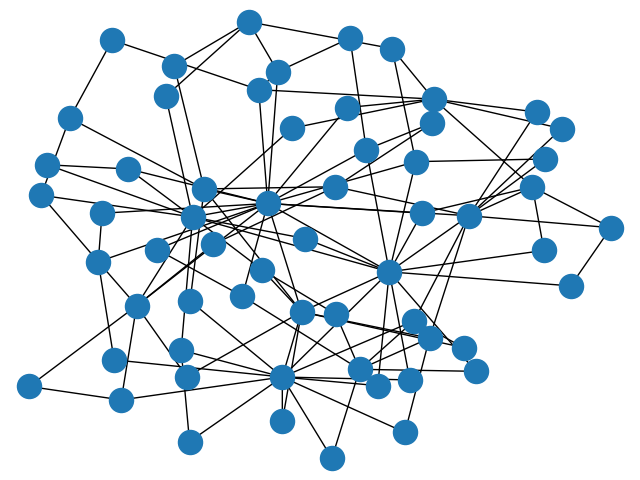}}
\caption{\label{figs:scalefreegraph} Exemplar Barab\'asi-Albert graphs with $k=1$ and $2$.}
\end{figure*}

To explore the role of the network topology in the resource apportionment
we consider a form of abstract complex graph for $\mathcal{M}$. 
While many options exist, including Erdos-Renyi and the Watts-Strogatz small world, we focus upon the Barab\'asi-Albert scale-free graph~\cite{albert2002statistical, barabasi2003emergence}, with examples in Fig.~\ref{figs:scalefreegraph}. 
Here, the probability of a node of degree $d$ follows a power-law
\begin{equation}\label{eqn:barbasi-centralities}
    p(d) \sim d^{-\gamma}
\end{equation}
with power exponent $\gamma\approx 3$. We will refer to Barab\'asi-Albert graphs with a suffix ($k$) denoting the degree of nodes
added in each step of the generating algorithm \cite{albert2002statistical}. 
Thus we use the notation $BA(k)$, where $k>1$ generates less sparse graphs with more paths or loops.

Because of its characteristic
hub-and-periphery structure, the scale-free graph has been argued to be demonstrated applicable to a broad range of human dynamics~\cite{song2010modelling, aymanns2017fake, shah2020finding}. 
For human organisations, what recommends this choice is the prevalence of hierarchy (where the hub is characteristically the apex of the hierarchy), but with a complexity richer than a simple tree-graph. However, 
for real organisations the purity of the
scale-free model is contested \cite{Li2009scalefree}. For our purposes,
the graph is simply a stylised choice which distinguishes the key elements of the topology, the hub and the periphery.
We also note that scale-free graphs allow for significant sizes (eg. number of users of the internet). The choice we explore here is `relatively' small, but nonetheless consistent with a larger 
version of a team, $N=100$.

We consider initial resource distributed across the graph based on a Boltzmann distribution of the eigencentralities $\hat{c}_i$, in which
\begin{equation}\label{eqn:pop_dist}
    P_i(T) = P_t \frac{\exp(\hat{c}_i T)}{\sum_{\forall i} \exp(\hat{c}_i T)}\enspace,
\end{equation}
where $T$ is the resource temperature, and $P_t$ is the total resource for a team. At zero temperature all resources are distributed evenly. Increasingly positive temperatures bias the resource distribution towards nodes with higher eigencentralities. Negative temperatures biases towards nodes with lower eigencentralities. We use eigencentralities $\hat{c}$ rather than centralities (described in Eq.~\eqref{eqn:barbasi-centralities}) due to their ability to better capture how a node is situated within the broader graph, rather than just its neighbours.

\section{Game Theory Solution Concepts}\label{sec:numerics}

Solutions to coupled nonlinear differential equations like the coupled system of Eqs.~\eqref{e:nbkl_positioning} and \eqref{eqn:nbkl_populations} are readily achievable through standard numerical methods. However, while such solutions are possible for any $\Phi$, to properly understand organisational dynamics it is important to understand how a team should position itself within $\Phi$ if it is playing rationally---a concept that can only be studied through game theory. With its inherent adversarial nature, our solution concept assumes each player is playing rationally to maximise its utility.

We therefore advance a previously developed framework \cite{cullen2022, demazy2018game}, in which the (non-networked) BKL engagement can be classified as a two-player, zero-sum, strategic game, where each player is a rational and strategic decision maker.
Though the strategy choices may be
across numerous variables of the model (resource number, the topology, or the couplings) here we select 
the frustration parameter as reflecting
the deliberate choice by teams how far ahead of the adversary decision `OODA' cycle~\cite{KALLONIATIS201621} they are attempting to operate {\it at different points in time}. Again, nonlinear dynamics may deny them their selected strategy. 
Within this context, for given sets of networked teams of agents $\mathcal{R}, \mathcal{B}$,
the frustration $\Phi$ is parameterised as
\begin{equation}
    \Phi_{ij}(t) = \begin{cases}
        \phi_k & \text{ if } i,j \in \mathcal{B}\\
        \psi_k & \text{ if } i,j \in \mathcal{R} \\
        0 & \text{ otherwise },        
    \end{cases}
\end{equation}
where the components $\phi_k \in [0, \pi]$ and $\psi_k \in [0, \pi]$ represent the phase offsets employed by each player at discrete time-steps, $t \in [k \delta_t, (k+1) \delta_t]$ for $k \in [0, \frac{T_f}{\delta_t}]$. Restricting $\phi$ and $\psi$ to changing at fixed temporal points facilitates computational tractability and deployment of exact game-theoretic solvers. Moreover, as the Networked BKL dynamics broadly exhibit exponential resource decay, introducing a finite game horizon $t \in [0, T]$ ensures that computing resources are focused upon the portions of the game which impact the final resource distribution.

At finite time horizon $T_f$, the game concludes. The resulting end-state of the NBKL system
quantifies the game outcome for the players who each optimise for
\begin{align} \label{e:staticbklutils}
U_{B}(S_{B}, S_{R}) = P_{B}(S_{B}, S_{R}) - P_{R}(S_{B}, S_{R}), \\
U_{R}(S_{B}, S_{R}) = P_{R}(S_{B}, S_{R}) - P_{B}(S_{B}, S_{R}),
\end{align}
where $P_{R}$ and $P_{B}$ are the aggregate populations of each team, which depend upon player strategies/actions
via the NBKL dynamics, and $S_B$ and $S_R$ represent the vectors of strategy choices across all decision points---the values of $\phi_k$ for $\mathcal{B}$ and $\psi_k$ for $\mathcal{R}$.

Each player chooses their respective strategy vector such that their behaviour follows the Nash-Equilibria (NE), which is the set of player strategies (and utilities) where no player gains deviating from their strategy when all other players also follow their own NE. This corresponds to a fixed point at the intersection of players' best responses (see Definition 3.22 of Ba\c{s}ar \cite{basargame}). However, pure-strategy NE may not exist in games like this. It is then natural to consider the \textit{security strategies} of players, which ensure a minimum performance. Also known as min-max and max-min strategies, these strategies allow each player to establish a worst-case bound on minimum outcome~\cite{gtessentialsbook}. Due to both the low-likelihood of repeated replay for NBKL scenarios and the computational cost of solving the underlying systems at large scales, we focus on pure strategies in contrast to probabilistic mixed strategies.

\begin{figure}[htbp!]
\begin{center}
\includegraphics[width=0.7\linewidth]{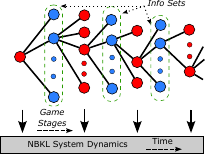} 
\end{center}
\caption{A game tree depicting discrete decisions of red and blue players (game stages) and the evolution of the NBKL dynamics. The information-sets apply across all actions so both players must act concurrently at each stage of the game.}
\label{fig:nbkltree}
\end{figure}

Constructing game theoretic solutions is made possible through \textit{Nash Dominant Game Pruning}~\cite{cullen2022}, which explores the game recursively, with utilities back-propagated up the game tree from the leaves. In contrast to a search over the full game tree, the Nash Dominant approach identifies action pairs $(\phi_k, \psi_k)$ that are strategically dominated---that will not correspond to the equilibrium state---and truncates subsequent exploration over the game tree. This truncation occurs by identifying if any row or column would not be picked by its corresponding player because a utility along that axis is smaller/larger than the minimum/maximum observed value along a row or column. If this is observed, any sub-games that stem from decisions along an incompletely resolved row or column can be truncated.

\section{Results}\label{sec:results} %

We now numerically solve NBKL games, considering scenarios
where each team uses either its highest or lowest centrality nodes 
at the adversarial surface.
Each team can choose from $4$ distinct values of frustration parameter at $4$ decision points in time, resulting in a game tree with $65,536$ leaf nodes. To ensure that these games are both deterministic and balanced both teams use the same fixed Barab\'asi-Albert graph of $100$ nodes. 

To examine the role of centrality in game outcomes and the decision making process, we vary across temperature values 
---following Eq.~\eqref{eqn:pop_dist}---to control initial distribution of resources across the nodes.
We consider different scenarios for the adversarial surface:
both sides using the three highest centrality nodes as their adversarial surface, the asymmetric
three Blue highest-vs-three Red lowest case, and both using the three lowest. Within the
adversarial surface we
consider when each three nodes
are individually arrayed against
one of the other side, ``1-vs-1'',
and where the three form a complete internal graph against the
opposition three, ``3-vs-3''.
Lastly, we analyse by temperature the degree to which players reposition their actions, namely dynamically adjust the frustration $\Phi$ -- how much they
adjust attempting to be ahead of the decision-cycle of the adversary in the Nash Equilibrium 
-- over the game dynamics.

\subsection{High vs High Centrality}

\begin{figure*}[htbp!]
   \subfloat[]{%
      \includegraphics[height=5cm]{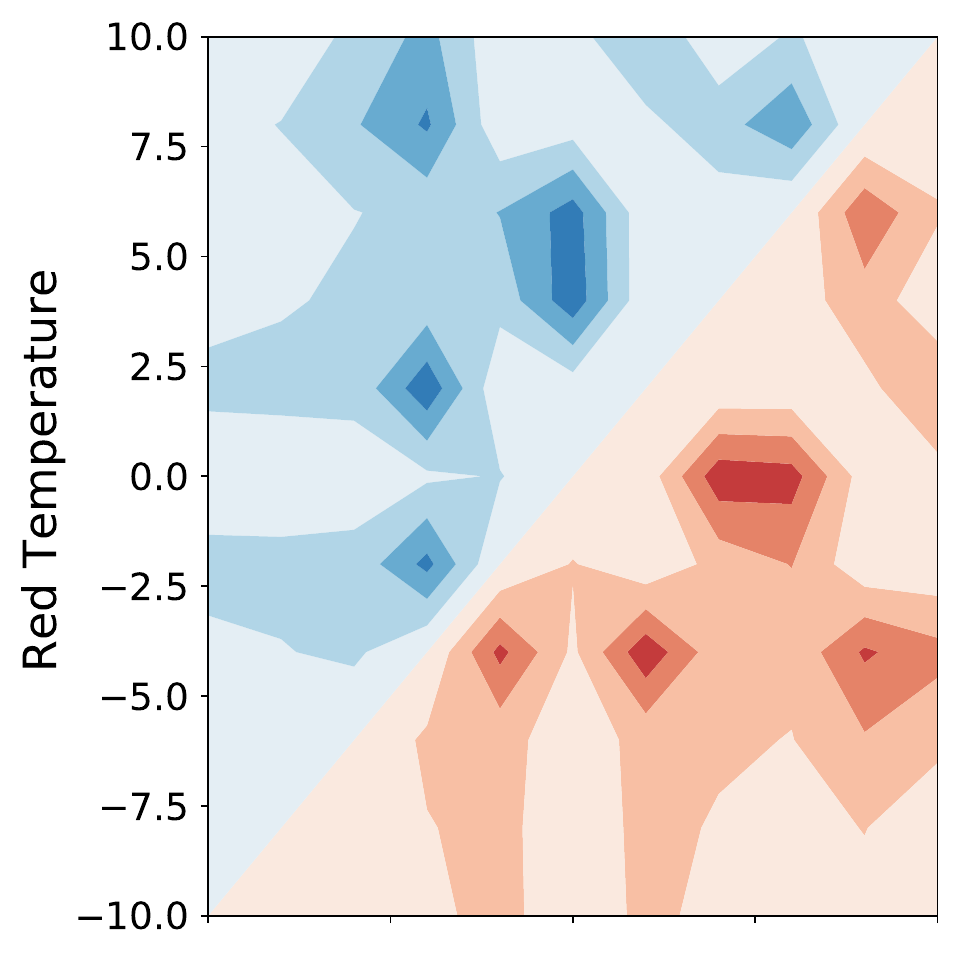}}
   \subfloat[]{%
      \includegraphics[height=5cm]{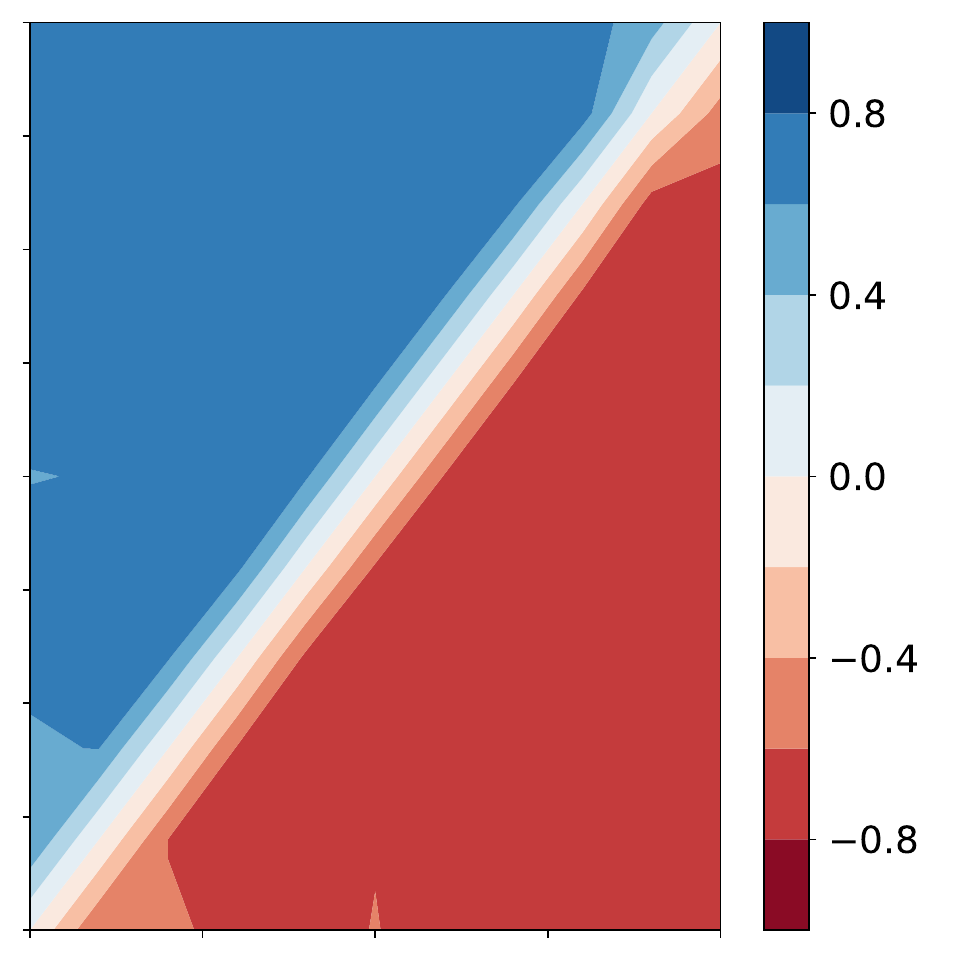}}\\
   \subfloat[]{%
      \includegraphics[height=5cm]{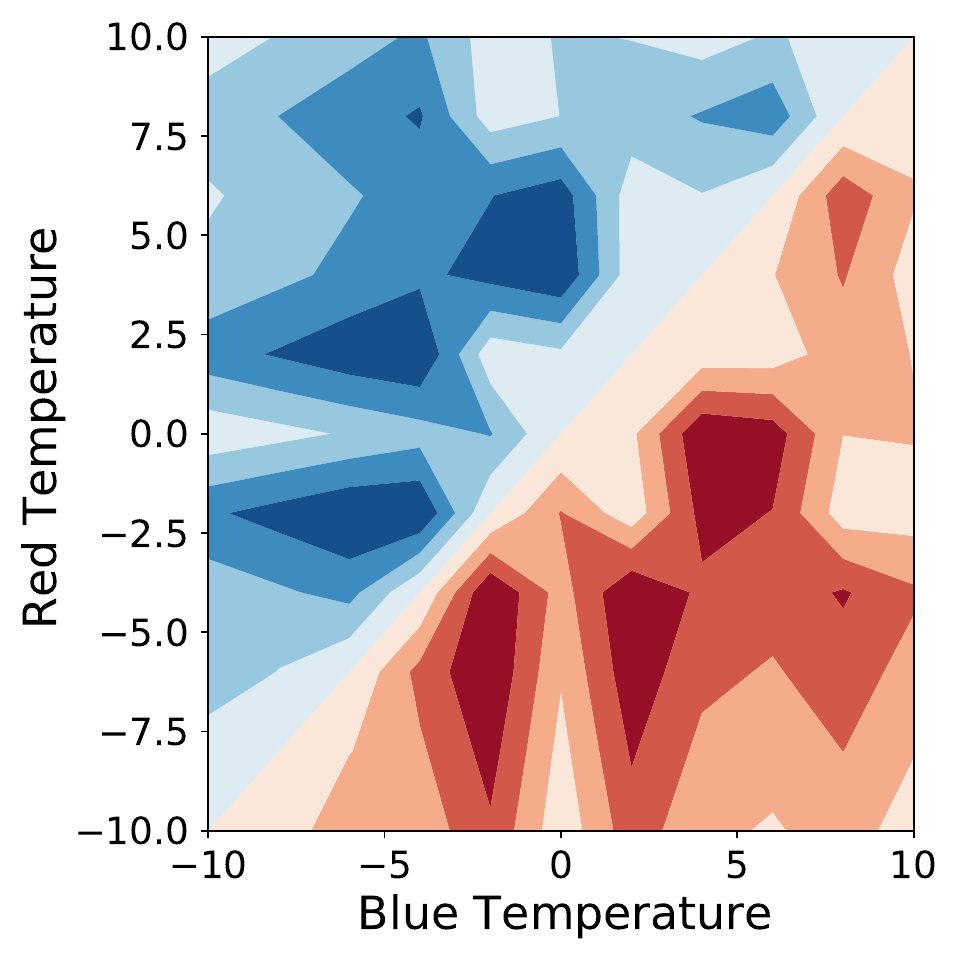}}
   \subfloat[]{%
      \includegraphics[height=5cm]{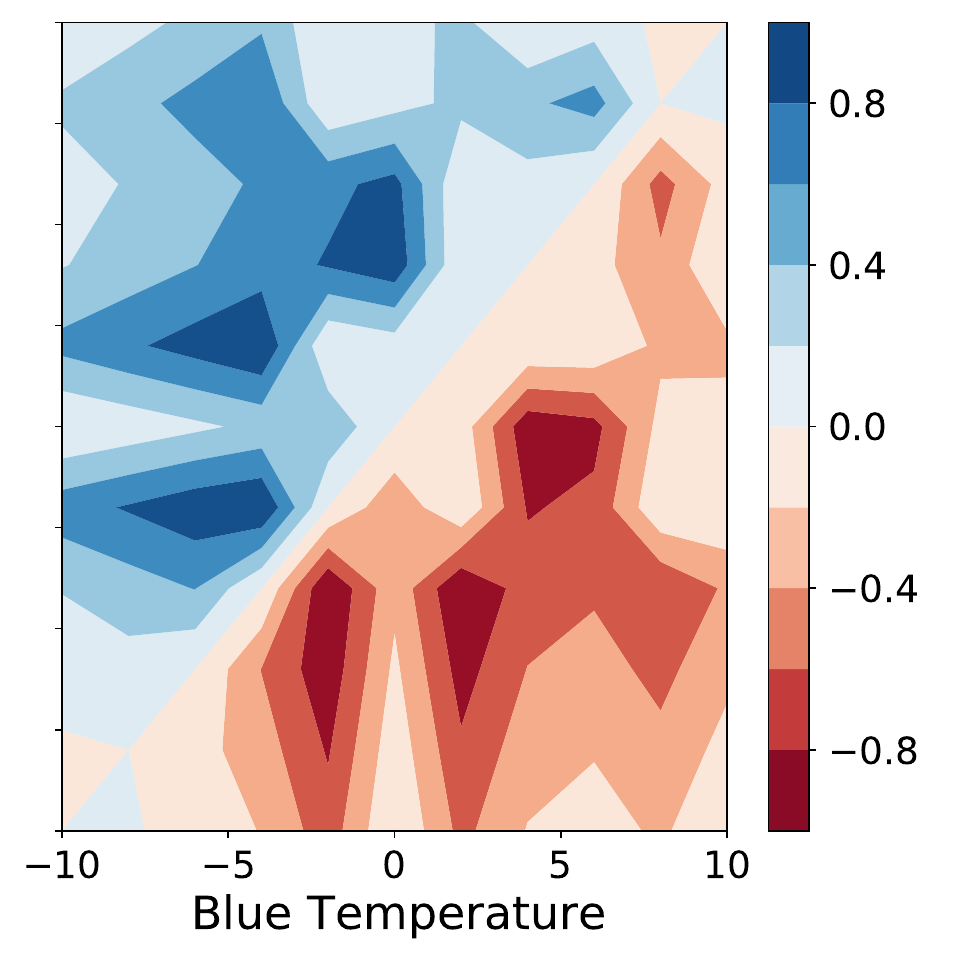}}         
\caption{\label{figs:HvH} Security Strategy Utilities (from the perspective of the Blue player) as a function of initial resource concentration temperature, where both players engage using their $3$ highest centrality nodes. The first and second rows respectively represent $BA(1)$ and $BA(2)$ graphs, while the first and second columns represent $3$ individual $1$-vs-$1$ and a complete $3$-vs-$3$ conflict respectively.
}
\end{figure*}

We begin by focusing upon the case where both players use their highest centrality nodes to compete against one another, the results of which can be seen in Fig.~\ref{figs:HvH}. Each point in the heatmap represents the utility at a Nash equilibrium for a specific variation by both teams of their frustration parameters at discrete points in time, to achieve decision advantage.  
The top row represents the $BA(1)$
graph, bottom row $BA(2)$; left column is 1-vs-1, and right 
3-vs-3.

Across all these we see a characteristic diagonal
axis of stalemate, with Blue dominating above and Red below the diagonal. Thus, a team prevails when it concentrates resources at the low-centrality nodes of the graph if the other side elects to concentrate at high-centrality. Thus Nash dominance is achieved if concentration is {\it away} from the nodes being used for adversarial interactions; resources must be initially in
reserve at the periphery if the hubs are the focus of competition. 

However, while the broad morphology of the diagonal stalemate is ubiquitous, the finer detail 
depends upon the internal connectivity within the graph and adversarial surface. While both $BA(2)$ graphs share similar behaviours, the $3$-vs-$3$ conflict scenario deviates around the diagonal when both players temperatures at the extreme top corner. The sign flip in the outcomes is numerically small, but suggests that more interlinked team structures offer
opportunity for success with
an ``all-in'' at the point of the
fight. 

The $BA(1)$ graphs show contrasting structures between the $1$-vs-$1$ and $3$-vs-$3$ scenarios. The latter differs
from all three cases here in its uniformity across temperature values.

\begin{figure*}[htbp!]
   \subfloat[]{%
      \includegraphics[height=5cm]{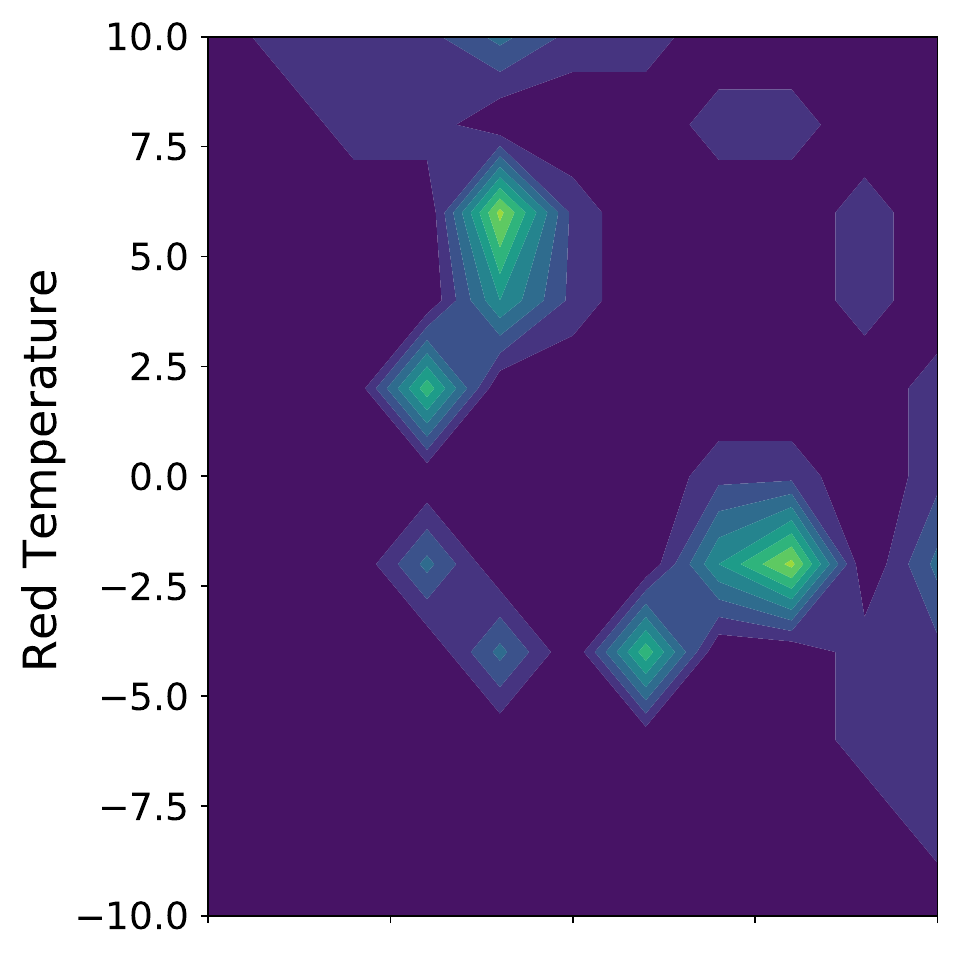}}
   \subfloat[]{%
      \includegraphics[height=5cm]{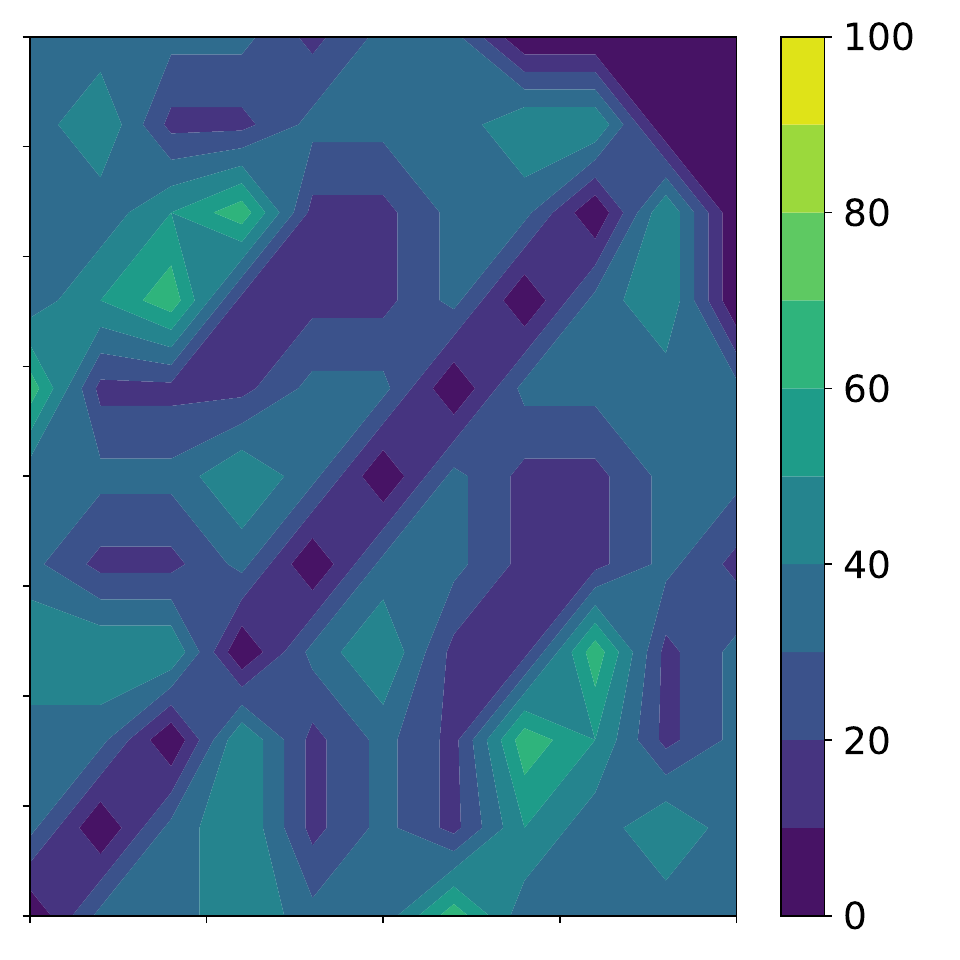}}\\
   \subfloat[]{%
      \includegraphics[height=5cm]{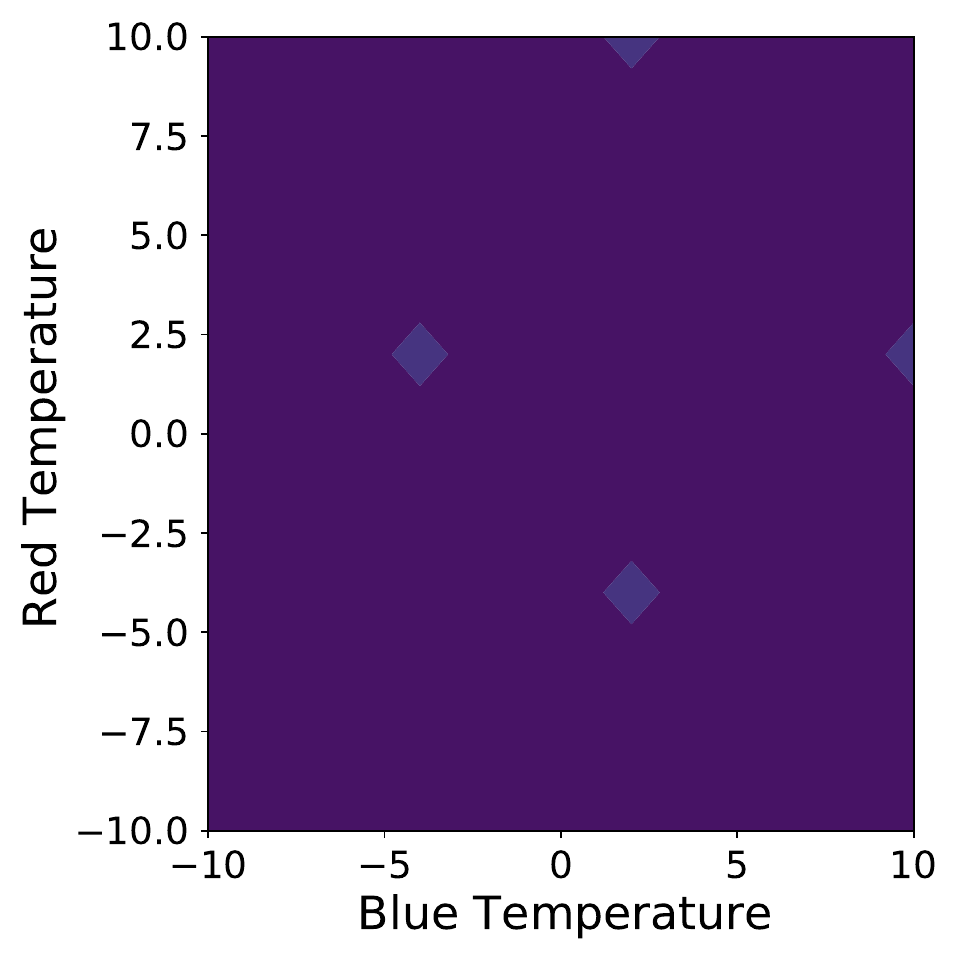}}
   \subfloat[]{%
      \includegraphics[height=5cm]{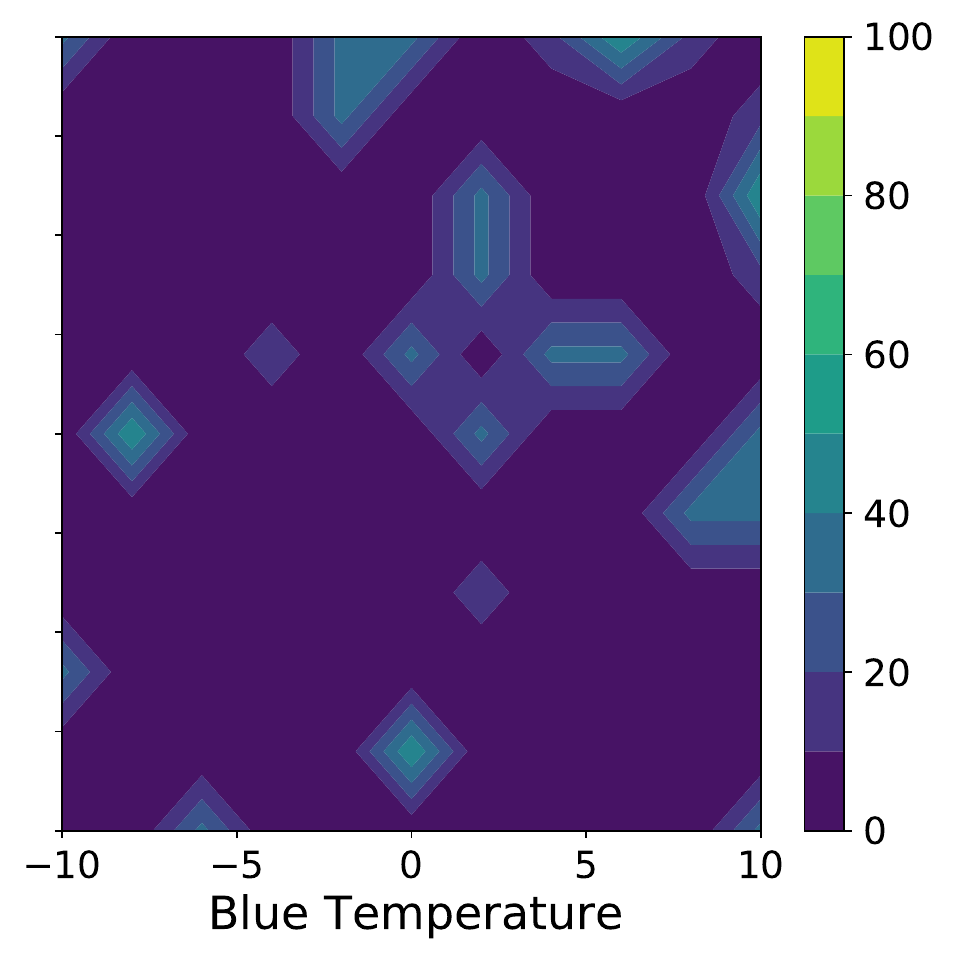}}         
\caption{\label{figs:HvH-Agility} Percentage ratio of repositioning actions taken by both players during the Nash Equilibrium policy, relative to the maximum possible repositioning actions, consistent with Fig.~\ref{figs:HvH}.
}
\end{figure*}

These structural differences are repeated in Fig.~\ref{figs:HvH-Agility}, 
showing the proportion of
repositioning in $\Phi$. 
The stalemate diagonal
is evident, and where the utility
heatmaps show variability across
temperature, we observe less
repositioning in $\Phi$. In contrast
the more uniform result in utilities for $BA(1)$ and 3-vs-3
coincides with more
repositioning of $\Phi$ during the dynamics. 
Overall, this suggests that for sparser graphs and higher connectivity
in the adversarial surface greater agility in decision-state
enables more effective manoeuvre of reserves into the competition.
In the remaining discussion we
will therefore identify
high repositioning of $\Phi$
with agility.

\subsection{High vs Low Centrality}

\begin{figure*}[htbp!]
   \subfloat[]{%
      \includegraphics[height=5cm]{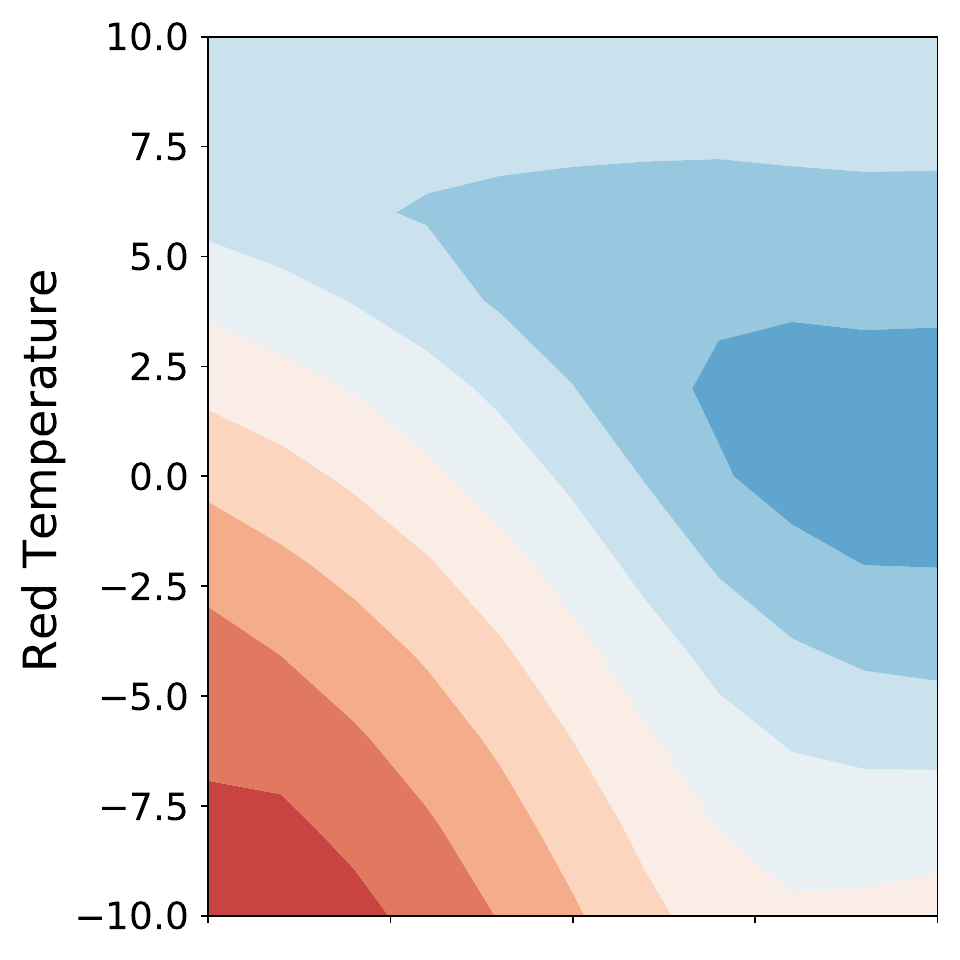}}
   \subfloat[]{%
      \includegraphics[height=5cm]{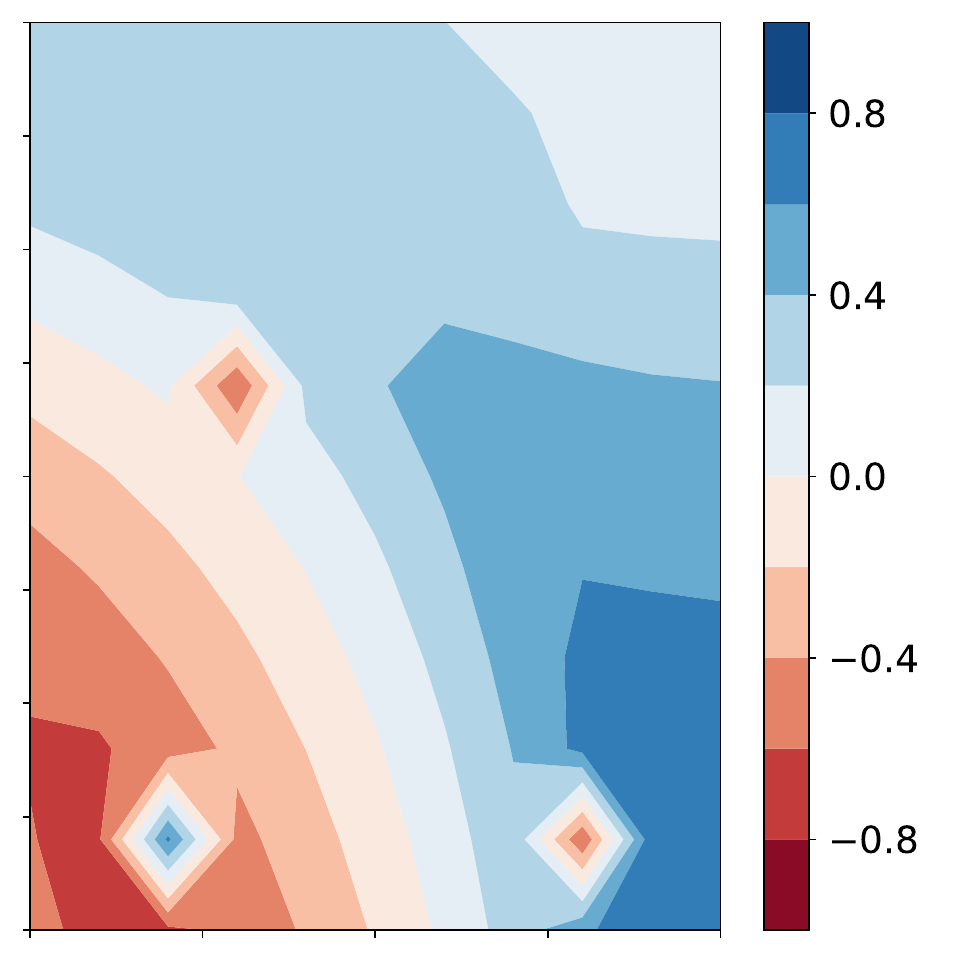}}\\
   \subfloat[]{%
      \includegraphics[height=5cm]{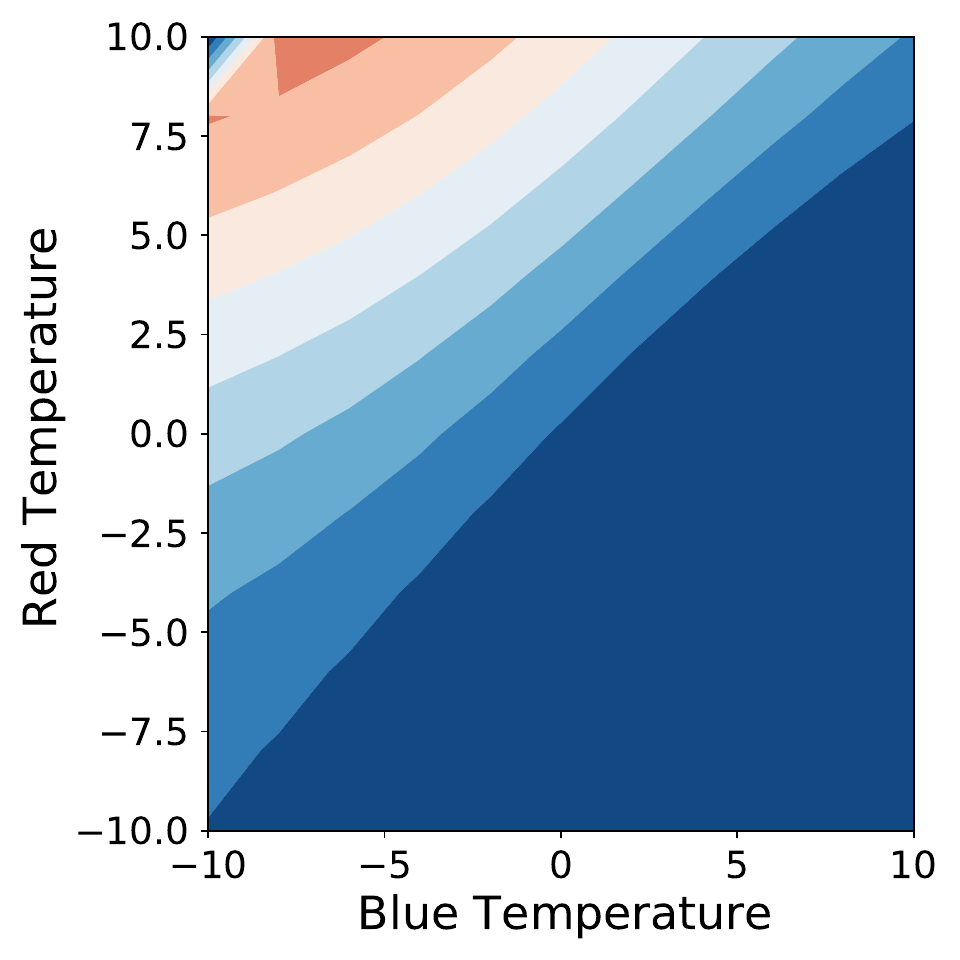}}
   \subfloat[]{%
      \includegraphics[height=5cm]{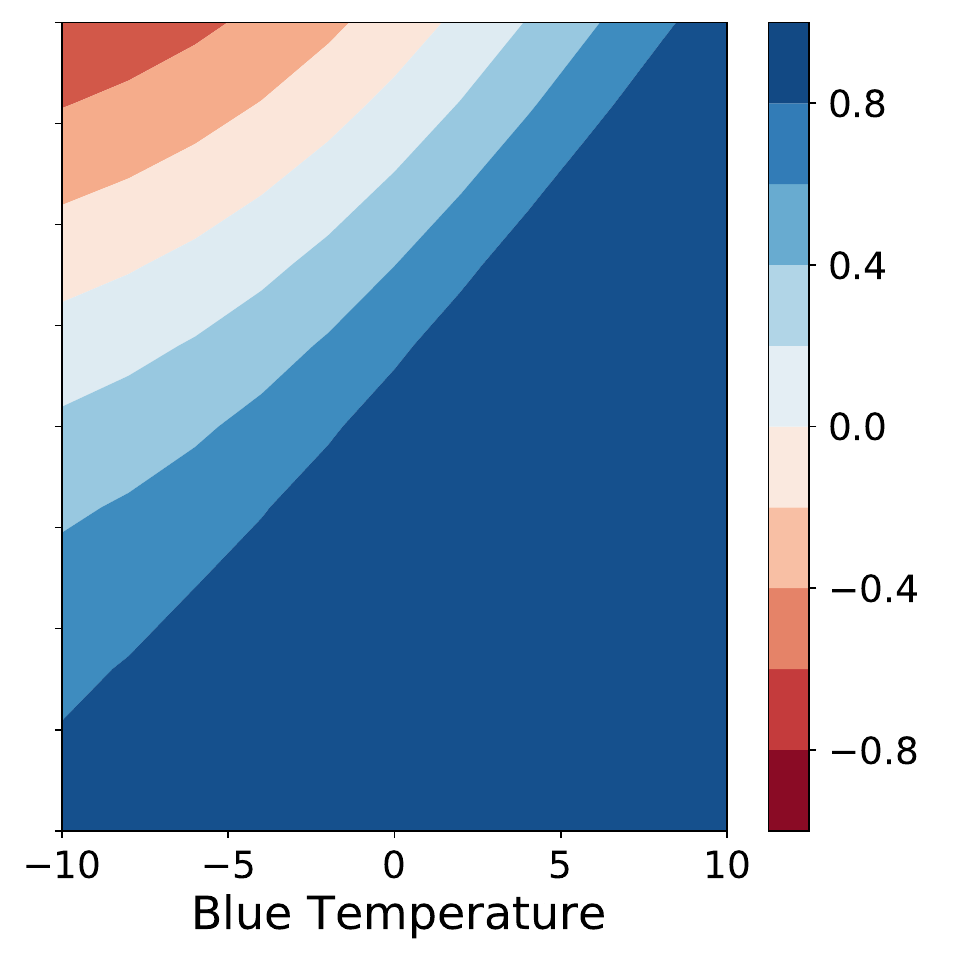}}         
\caption{\label{figs:HvL} Security Strategy Utilities (from the perspective of the Blue player) as a function of initial resource concentration temperature, where the Blue player is using their highest centrality nodes against the lowest centrality Red nodes. The panels repeat the arrangement
of Fig.\ref{figs:HvH}
}
\end{figure*}

In the asymmetric case of Blue high- against Red low-centrality the diagonal
symmetry disappears, as seen in Fig.~\ref{figs:HvL}. Nash dominance is consistently achieved by the high-centrality team when it concentrates resources at high-centrality nodes, namely at the adversarial surface. For the $BA(1)$ graph, the low-centrality team should also centralise its resources at its
low-centrality nodes on the adversarial surface. Neither side gains from reserving resource, a consequence of the structural asymmetry between the two sides. In particular, we see that however much Red seeks to concentrate at low-centrality nodes, Blue can always initially concentrate more at its high-centrality nodes. Thus, Blue seeks to overwhelm through structural superiority, and Red cannot afford to hold back because of structural inferiority. 
For asymmetry, committing all-in is important to exploit
resource advantage or mitigate it when lacking superior numbers at the point of the fight. The only
exception to this is the occurrence
of discrete poles in the $BA(1)$
3-vs-3 case, suggesting local
deviations within the random structure for intermediate
initial resource distributions.
These morphological sensitivities persist with different random seed for graph generation, and the number of nodes within each team's graph. This indicates that these behaviours are a property of 
variability within the bulk characteristics of the graph, rather than just its finer structure. 

\begin{figure*}[htbp!]
   \subfloat[]{%
      \includegraphics[height=5cm]{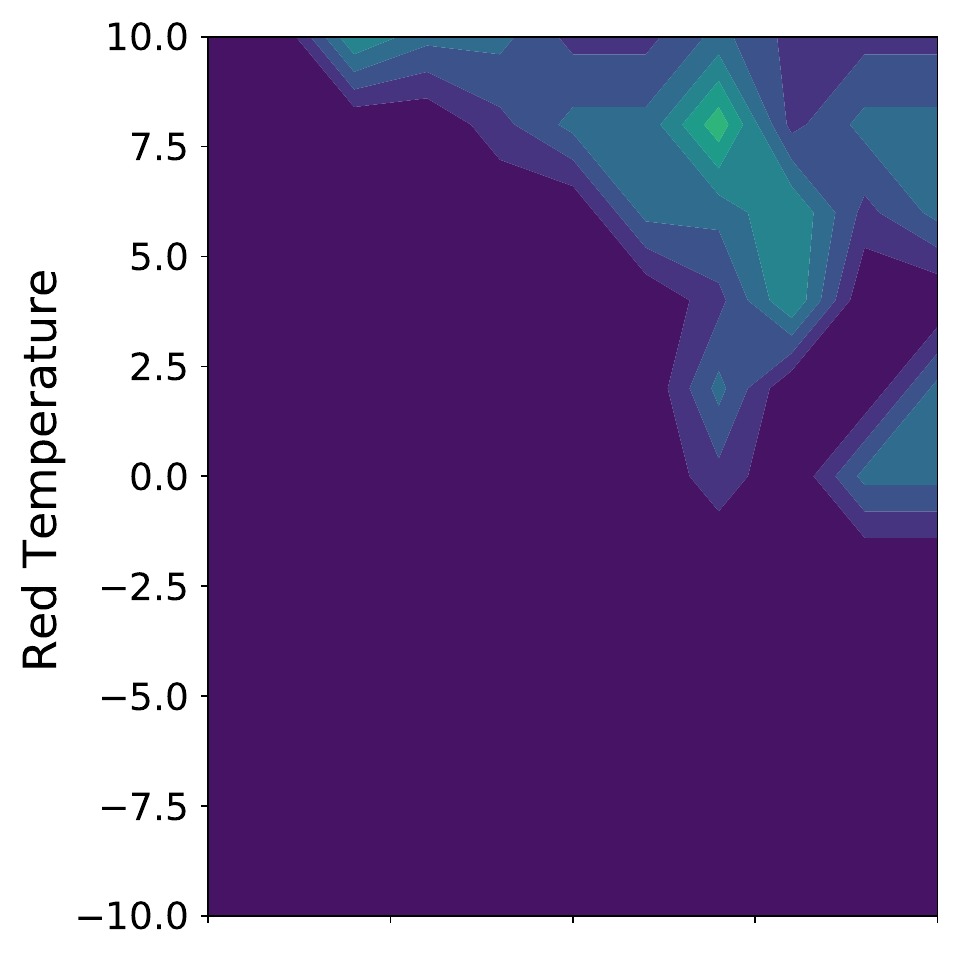}}
   \subfloat[]{%
      \includegraphics[height=5cm]{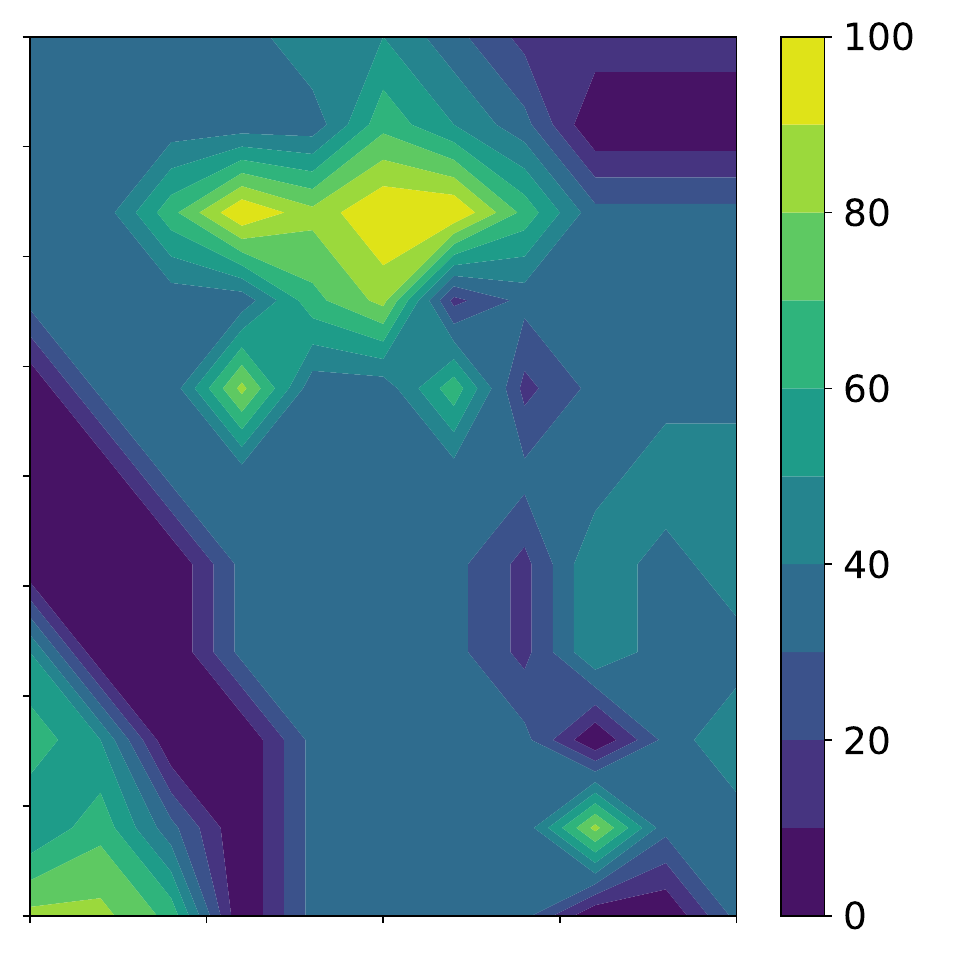}}\\
   \subfloat[]{%
      \includegraphics[height=5cm]{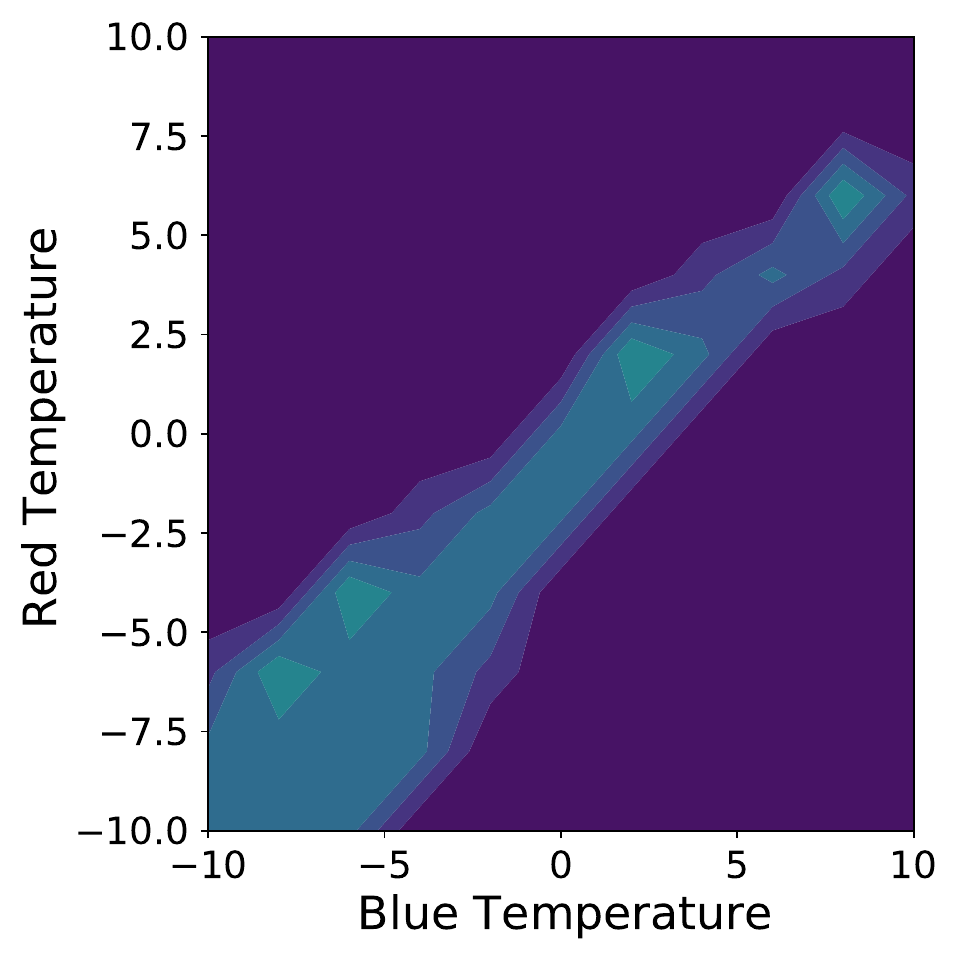}}
   \subfloat[]{%
      \includegraphics[height=5cm]{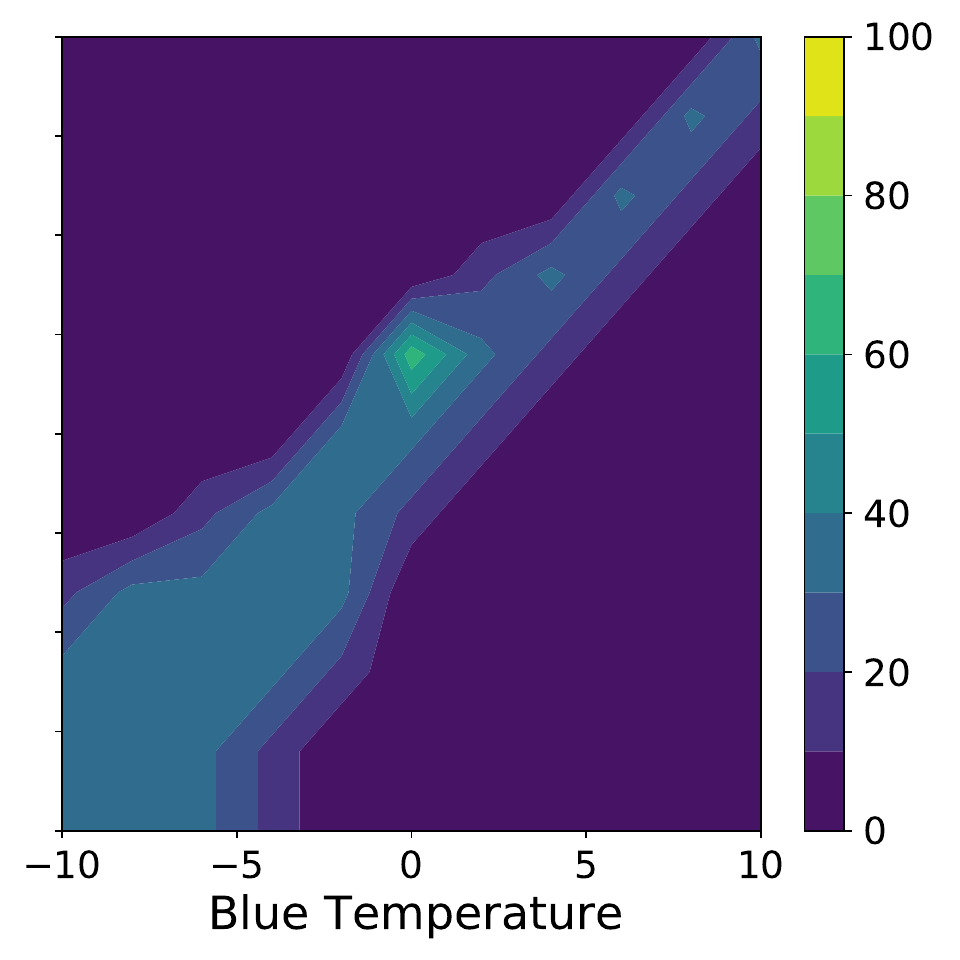}}         
\caption{\label{figs:HvL-Agility} Percentage ratio of repositioning actions taken by both players during the Nash Equilibrium policy, relative to the maximum possible repositioning actions, consistent with Fig.~\ref{figs:HvL}.
}
\end{figure*}

In contrast to the High vs High (and, as we will establish forthwith the Low vs Low) scenario, the most highly polarised results are not correlated with the highest agility behaviours shown in Fig.~\ref{figs:HvL-Agility}. This, alongside the broad similarities between the $3$ $1$-vs-$1$ and $3$-vs-$3$ conflict suggests that the structural asymmetry of the High vs Low scenario is a structural factor that cannot be overcome by the agility of the players. 

\subsection{Low vs Low Centrality}

\begin{figure*}[htbp!]
   \subfloat[]{%
      \includegraphics[height=5cm]{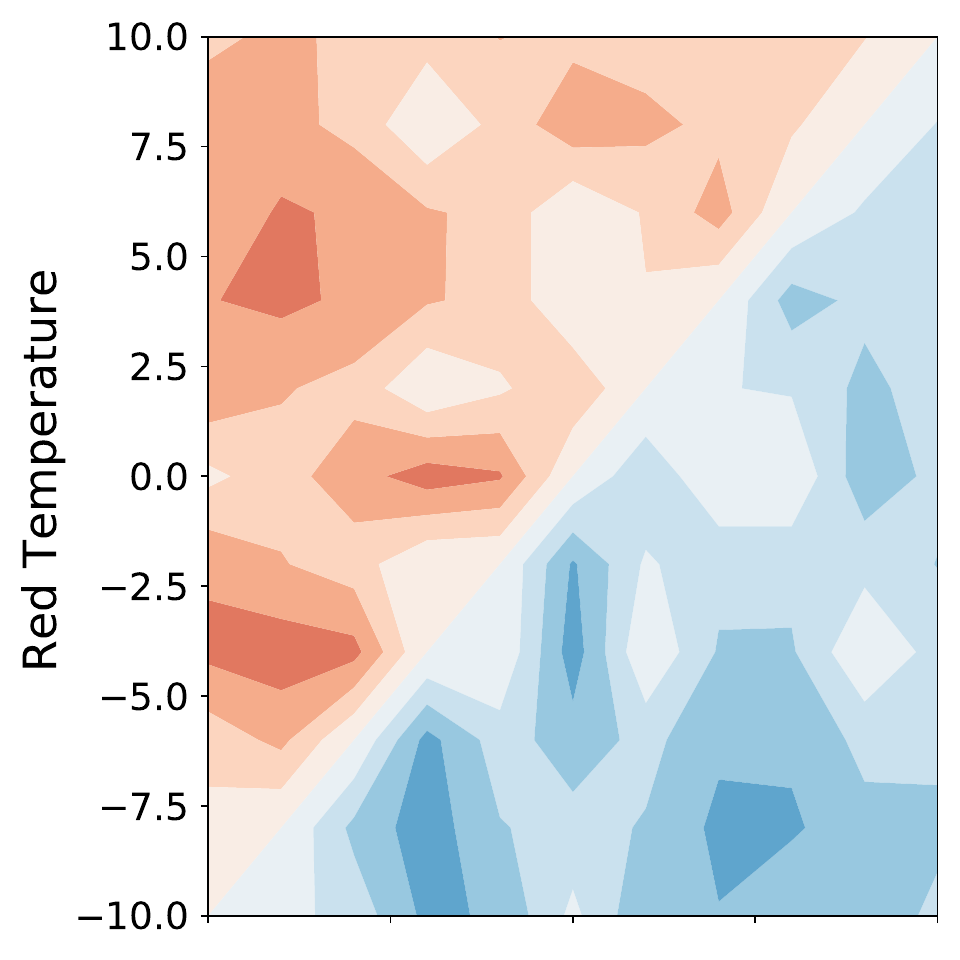}}
   \subfloat[]{%
      \includegraphics[height=5cm]{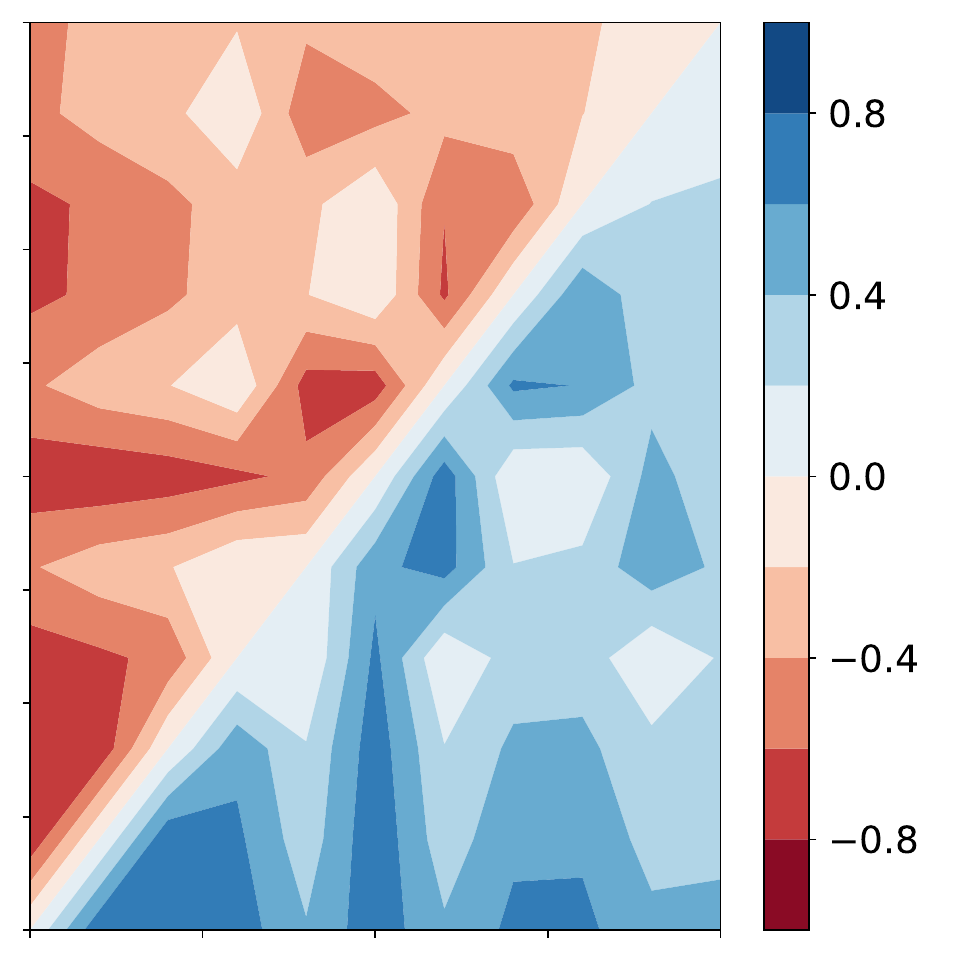}}\\
   \subfloat[]{%
      \includegraphics[height=5cm]{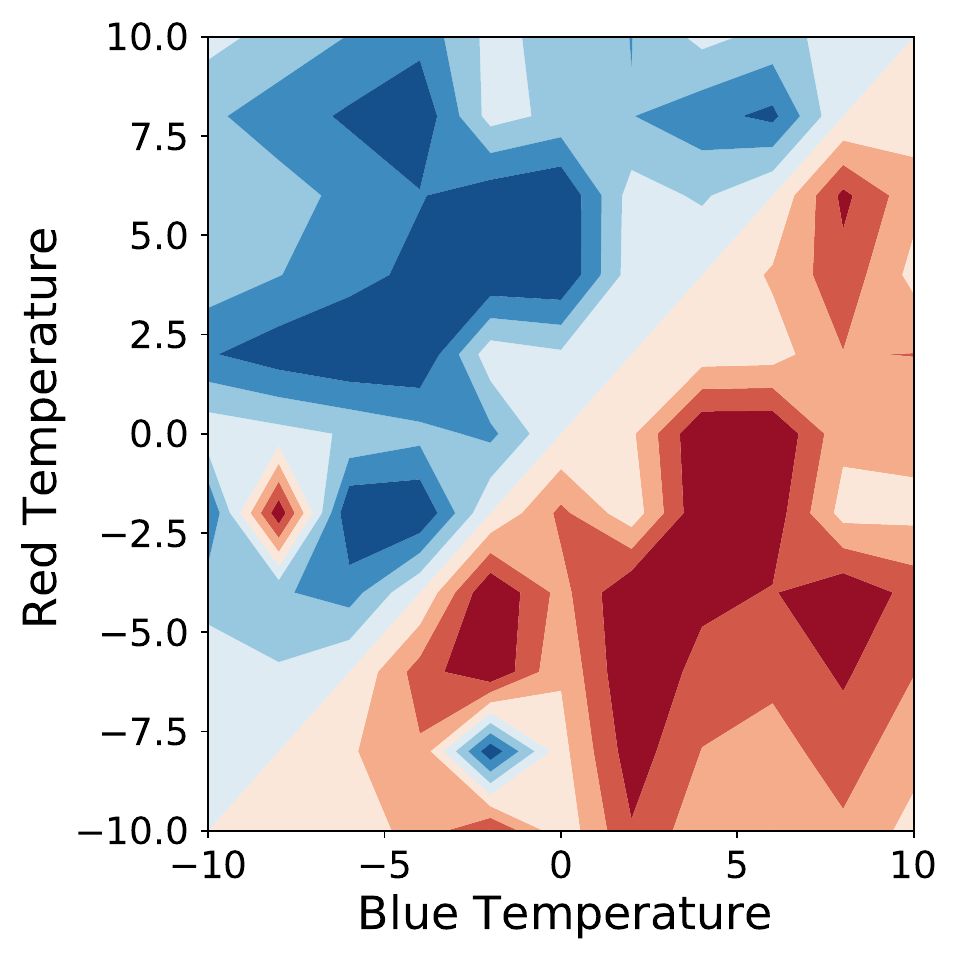}}
   \subfloat[]{%
      \includegraphics[height=5cm]{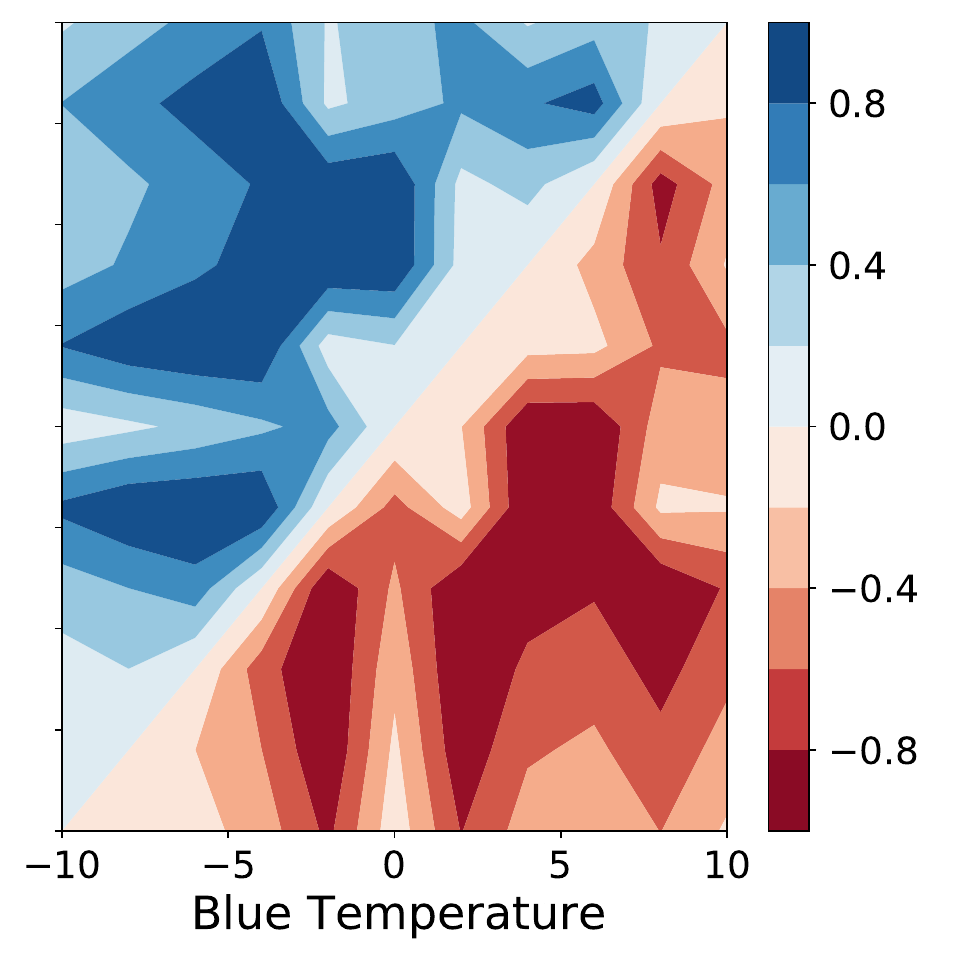}}         
\caption{\label{figs:LvL} Security Strategy Utilities (from the perspective of the Blue player) as a function of initial resource concentration temperature, where both players engage in conflict using their $3$ lowest centrality nodes.The panels repeat the arrangement of 
Fig.~\ref{figs:HvH}.
}
\end{figure*}

When both players use their $3$ low-centrality nodes (graph periphery) for the adversarial surface, Fig.~\ref{figs:LvL}, 
we see yet another variation in behaviours. For $BA(1)$, results are inverted across the diagonal, relative to the High vs High scenario. Here then 
Nash dominance is achieved by the low-centrality team when it concentrates resources at the high centrality nodes, now away from the adversarial surface. Effectively, the limited paths to and narrowness of the bottleneck at the
point of engagement requires that reserves be distributed. 
In the $BA(2)$ case we 
are back to the High-High case, with players incentivised to draw all their resources to the adversarial surface; more paths
allow closer concentration to the engagement. In
the 1-vs-1 case we see another
example of isolated poles, a product of the inherently nonlinear chaotic dynamics 
with greater connectivity, again
repeated with different random seed. 

\begin{figure*}[htbp!]
   \subfloat[]{%
      \includegraphics[height=5cm]{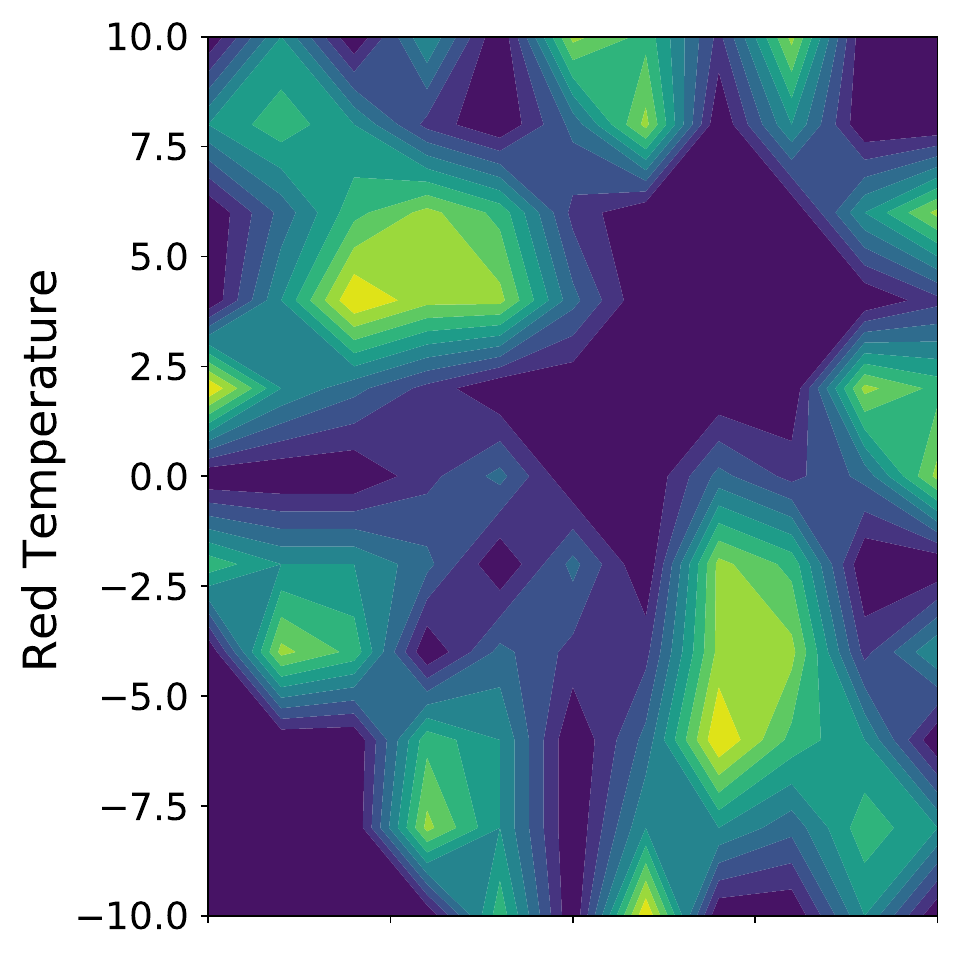}}
   \subfloat[]{%
      \includegraphics[height=5cm]{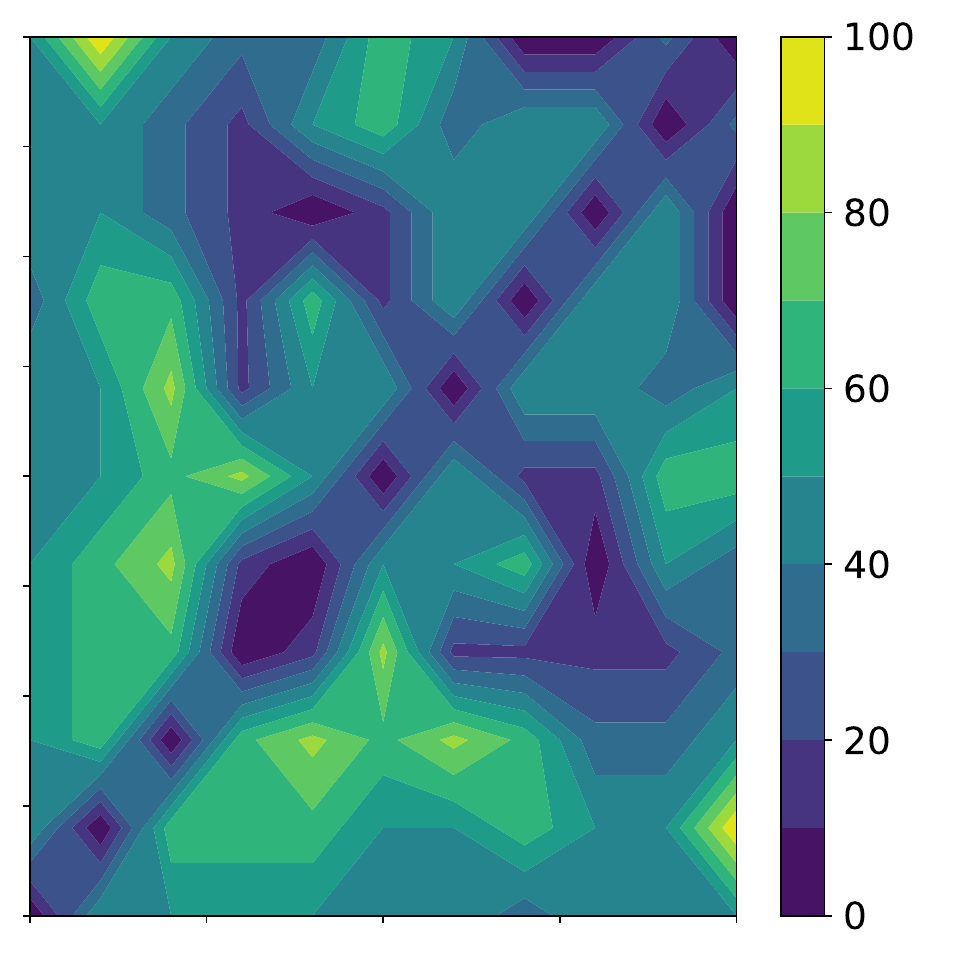}}\\
   \subfloat[]{%
      \includegraphics[height=5cm]{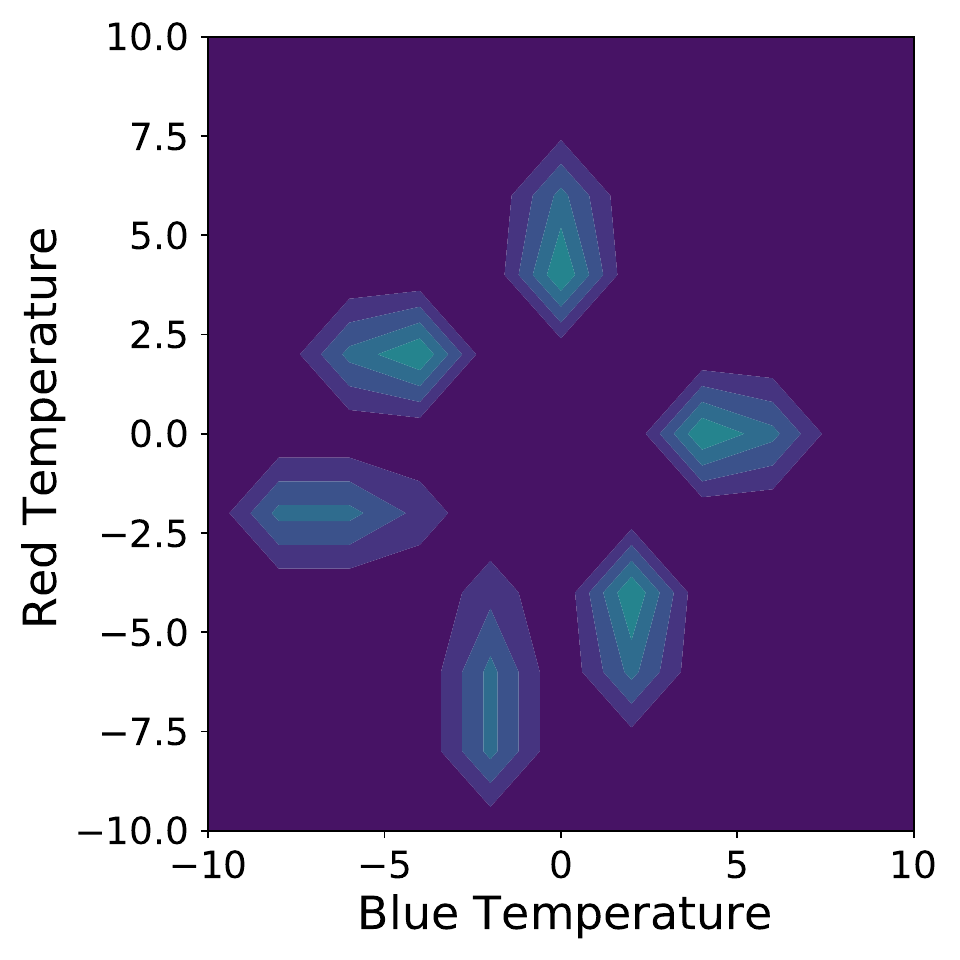}}
   \subfloat[]{%
      \includegraphics[height=5cm]{Figures/100/Agility/BA2-3v3-hh_BR_c.pdf}}         
\caption{\label{figs:LvL-Agility} Percentage ratio of repositioning actions taken by both players during the Nash Equilibrium policy, relative to the maximum possible repositioning actions, consistent with Fig.~\ref{figs:LvL}.
}
\end{figure*}

Inspecting the agility across these cases, Fig.~\ref{figs:LvL-Agility} shows broad similarities to Fig.~\ref{figs:HvH-Agility}, in that the extrema of the Nash Equilibrium utilities correlate to the parts of the temperature space in which the players are more agile. The exception is the $BA(1)$ 1-vs-1 result where the player repositioning 
is consistent with higher utility.
As the most topologically constrained case, this shows that
agility in seeking
decision-advantage provides for
higher utility.

\section{Conclusion}\label{sec:conclusion}
Analysis of adversarial team interactions modelled as
the Networked Boyd-Kuramoto-Lanchester equations through a game theoretic lens yields insights into the role
of network topology in 'victory' or
'defeat'. By testing scale-free graphs with
a characteristic hub-periphery topology, we were able to study the impact of structural changes in terms of the interlinking of teams, and their resource distributions.
In the case of sparsely connected teams, 
we find that initial
resources should be held in reserve,
unless there is an asymmetry in the
connectivity of agents at the focus
of the engagement. If the hub is that point against the other team's periphery, then the former team
must exploit that advantage. Concomitantly, for the team engaging with its periphery concentrating resource elsewhere than at those nodes makes a bad situation worse. 
However, as the density of the connections within the team graph increases, there is a need to transition towards greater resource centralisation
at the point of engagement except when
both teams employ their hub nodes at
the adversarial surface. 
These results represent Nash equilibria in decision-advantage in relation to
the other, modelled by the frustration, where the
nonlinearity of the 
Kuramoto-Lanchester model captures
the complexity of resource competition
in networked organisations.

We have seen that across the
scenarios there are cases where
agility in that decision-state
over time contributes to the
utility outcome. In essence,
structure dominates
agility is less consequential;
when structure is sparse then
agility becomes a significant 
mechanism for achieving success.
We have identified these outcomes
from a single graph instance; future work will systematically classify this over larger ensembles.

This work demonstrates that resource apportionment should be considered in parallel with 
structural designs of organisations 
locked in competition with an opponent and agile decision-making. That such insights are able to be gleaned from the combination of game theory and nonlinear dynamics indicates the value of such sociophysical modelling in a variety of contexts, including military strategy, business marketing strategy and cyber-security, to name but a few.  %

\newpage

\bibliography{nbklgame}%

\end{document}